\documentclass[
 aip,jcp,
 amsmath,amssymb,
 reprint
]{revtex4-2}

\usepackage{haimi}
\usepackage[T1]{fontenc}
\usepackage{mathptmx}
\usepackage{etoolbox}
\usepackage{physics}
\usepackage{nicefrac}
\usepackage{empheq}
\usepackage{xcolor}
\usepackage[normalem]{ulem}

\newcommand{\hnadd}[1]{{\color{BlueViolet} #1}}
\newcommand{\hncomment}[1]{}
\newcommand{\hndel}[1]{}
\newcommand{\gkccomment}[1]{}
\newcommand{\llcomment}[1]{}
\newcommand{\gpcomment}[1]{}

\makeatletter
\def\@email#1#2{%
 \endgroup
 \patchcmd{\titleblock@produce}
  {\frontmatter@RRAPformat}
  {\frontmatter@RRAPformat{\produce@RRAP{*#1\href{mailto:#2}{#2}}}\frontmatter@RRAPformat}
  {}{}
}%
\makeatother

\renewcommand{\spin}[1]{\sigma^{#1}}
\renewcommand{\Re}{\operatorname{Re}}

\begin{document}

%\preprint{AIP/123-QED}

%\title{Equilibrium Correlation Functions From Tensor Network Methods: The Case of the Spin-Boson Model}
\title{Correlation Functions From Tensor Network Influence Functionals: The Case of the Spin-Boson Model}
\author{Haimi Nguyen}
 %\email{hn2375@columbia.edu}
 \affiliation{Department of Chemistry, Columbia University, NY 10027, USA}
\author{Nathan Ng}
 \affiliation{Department of Chemistry, Columbia University, NY 10027, USA}
\author{Lachlan P.\ Lindoy}
\affiliation{National Physical Laboratory, Teddington, TW11 0LW, United Kingdom}
\author{Gunhee Park}
\affiliation{Division of Engineering and Applied Science, California Institute of Technology, Pasadena, California 91125, USA}
\author{Andrew J.\ Millis}
 \affiliation{Department of Physics, Columbia University, NY 10027, USA}
 \affiliation{Center for Computational Quantum Physics, Flatiron Institute, New York, New York 10010, USA}
\author{Garnet Kin-Lic Chan}
 \affiliation{Division of Chemistry and Chemical Engineering, California Institute of Technology, Pasadena, California 91125, USA}
\author{David R.\ Reichman}
 \email{drr2103@columbia.edu}
 \affiliation{Department of Chemistry, Columbia University, NY 10027, USA}

\date{\today}

\begin{abstract}
We investigate the application of matrix product state (MPS) representations of the influence functionals (IF) for the calculation of real-time equilibrium correlation functions in open quantum systems.
Focusing specifically on the unbiased spin-boson model, we explore the use of IF-MPSs for complex time propagation, as well as IF-MPSs for constructing correlation functions in the steady state.
We examine three different IF approaches: one based on the Kadanoff-Baym contour targeting correlation functions at all times, one based on a complex contour targeting the correlation function at a single time, and a steady state formulation which avoids imaginary or complex times, while providing access to correlation functions at all times. We show that within the IF language, the steady state formulation provides a powerful approach to evaluate equilibrium correlation functions.
\end{abstract}

\maketitle

\section{Introduction}

Understanding the dynamical behavior of open quantum systems remains one of the most challenging and exciting frontiers in condensed matter physics and quantum information theory. Central to the field are two-time correlation functions, which offer important information about the physical properties of a system and report on phenomena such as phase transitions, quantum entanglement, and the emergence of collective excitations~\cite{BreuerBook, Sachdev2011, WeissBook, ForsterBook}.
Within linear response theory, two-point equilibrium correlation functions describe the spectroscopic, transport, and chemical kinetic properties of open systems found in chemical and condensed matter physics~\cite{ForsterBook, NitzanBook, Kadanoff1963}.

A special and important example of a quantum many body system is that of an open quantum system with linear coupling to a quadratic bath. In this case,  
a salient object is the Feynman-Vernon influence functional~\cite{feynman2000theory}, which reweights the path integral of the system and captures all the effects of the dynamics from the bath. 
Recent studies~\cite{Banuls2009, Strathearn2018, Jorgensen2019, Ye2021, Lerose2021, Sonner2021, Thoenniss2023, NgPark2023, park2024tensor} have suggested that, under conditions related to the fast decay of temporal correlations, tensor networks such as matrix product states (MPS) provide an efficient representation of influence functionals.
This observation has been exemplified by the development of algorithms such as Time-Evolving Matrix Product Operators (TEMPO)~\cite{Strathearn2018} and related methodologies~\cite{Jorgensen2019, Gribben2022, FowlerWright2022, Cygorek2024}, which provide a new path for the simulation of open quantum systems. 

While the utility of tensor network influence functionals for simulating real-time quench dynamics of open quantum systems is by now well-demonstrated, 
there has been less work on computing the equilibrium dynamical correlation functions, which involve the non-trivial equilibrium state.
Computing correlation functions of general non-Markovian open quantum systems have only received attention very recently~\cite{Bose2023,Cygorek2024}~\footnote{The Markovian case constitutes a trivial limit for tensor network path integrals, and can be treated by other means~\cite{Wolff2020, Nathan2020, Carballeira2021}}.

In this work, we will focus on the application of tensor network influence functionals for the simulation of two particular types of correlation functions in systems linearly coupled to a quadratic non-Markovian bath.
Defining the equilibrium density $\rho_{\beta} = \exp(-\beta H)/Z$ and observables $A$ and $B$ such that $A(t) = e^{i H t} A e^{-i H t}$,
one has the equilibrium correlation function,
\begin{align}
    \label{eq:equilibrium-corr}
    G_{AB}(t) &= \Theta(t) \Tr(\rho_{\beta} A(t) B),
\end{align}
and its symmetrized relative,
\begin{align}
    \label{eq:symmetrized-corr}
    C_{AB}(t) &= \Theta(t) \Tr(\sqrt{\rho_{\beta}} A(t) \sqrt{\rho_{\beta}} B) \\
    &= \Theta(t) \Tr(\rho_{\beta} A(t-i \beta/2) B) \nonumber \\
    &= G_{AB}(t-i\beta /2). \nonumber
\end{align}
Here, $\Theta(t)$ is the Heaviside function.
It has long been appreciated that the symmetrized correlation function, $C_{AB}(t)$, contains the same physical content as $G_{AB}(t)$ while simultaneously containing features that render theoretical manipulations more transparent~\cite{Miller1983, Bonella2010a, Bonella2010b}.
In the time-domain, the two correlation functions are related by an analytic continuation to complex time, cf.\ Eqs.~\eqref{eq:equilibrium-corr} and \eqref{eq:symmetrized-corr}.
In Fourier space, the relation becomes
\begin{align}
    \label{eq:relation}
    \widetilde{C}_{AB}(\omega) &= \exp(-\beta \omega/2) \widetilde{G}_{AB}(\omega),
\end{align}
so that in principle, we should be free to choose whichever quantity is easier to calculate, since 
$G_{AB}(t > 0) = (1/2\pi) \int d\omega \, \widetilde{G}_{AB}(\omega) e^{-i \omega t}$.

While these correlation functions are formally related to each other, practically, there may be limitations in obtaining one from the other. For example, obtaining $G_{AB}(t)$ from $C_{AB}(t)$ through Eq.~\eqref{eq:relation} may suffer from exponential amplification of features at high frequencies.

Despite this connection, we note that there is independent interest in obtaining the symmetrized correlation functions.
Since the low frequency features between $G_{AB}(t)$ and $C_{AB}(t)$ are not greatly altered, this allows for the use of $C_{AB}(t)$ to investigate long time dynamics such as rate behavior.
In addition to its relevance for rate theory calculations, higher-order versions of the symmetrized correlator and its relatives are important quantities in other domains.
For example, the regularized out-of-time-order correlators (OTOCs), which are used to probe quantum chaos~\cite{Larkin1969, Aleiner2016, Maldacena2016, GarcaMata2023}, can be viewed as symmetrized correlators in a replicated space~\cite{Pappalardi2022}.
These correlators are subject to simple analyticity constraints~\cite{Maldacena2016}, which have been found to be important in characterizing the dynamics of operator complexity~\cite{Parker2019}.
Finally, we note that other studies have found it more computationally useful to compute the symmetrized correlator over the thermal one.  Liu {\it et al.}~\cite{Liu2022} have shown that the Monte Carlo sampling of imaginary-time symmetrized correlations provides a concrete route to real-time dynamical correlators.  Kobrin {\it et al.}~\cite{Kobrin2021} have shown that the symmetrized OTOC exhibits a more facile numerical route to extract the Lyapunov exponent in the Sachdev-Ye-Kitaev model.
Lastly, we note that this quantity has been studied using numerical path integral methods (with~\cite{Bose2023} and without~\cite{Topaler1993} tensor compression), hierarchical equations of motion~\cite{Song2015}, as well as the aforementioned open-chain path integral molecular dynamics~\cite{Liu2022}.

In addition to calculating these
two related quantities using tensor network influence functionals defined on complex time-contours, we will also present an approach to obtaining equilibrium correlation functions that only uses the influence functionals along the real-time axis. The basic idea is that if we prepare the bath at the appropriate temperature, the dynamics itself will thermalize the open system in the steady state, and this steady state is encoded in the tensor network representation of the real-time influence functionals.
We compare the advantages and disadvantages of these approaches to obtaining the equilibrium correlation function information.

This paper is organized as follows: 
in Sec.~\ref{sec:method} we first detail how equilibrium correlation functions can be computed from influence functionals describing propagations in imaginary- and real-time, as well as how they can be extracted from the steady state limit of influence functionals in real-time.
In Sec.~\ref{sec:results} we compare the computational costs associated with the different approaches and demonstrate cases where it is preferable to calculate equilibrium correlation functions from the steady state limit of real-time propagation as opposed to symmetrized correlation functions from complex-time propagation. 
We conclude in Sec.~\ref{sec:discussion} by discussing the utility of tensor network techniques for calculating correlation functions in open quantum systems.

\section{\label{sec:method}Methods}

    \begin{figure}
        \centering
        \includegraphics[width=\columnwidth]{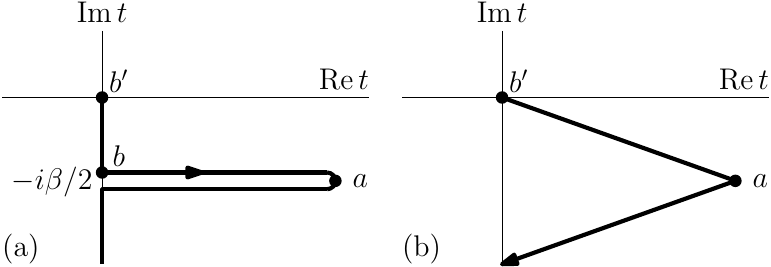}
        \caption{Schematic depiction of the contours relevant for the calculation of equilibrium correlation functions \textbf{(a)} Kadanoff-Baym-like (KB) contour, separating imaginary-time propagations from real-time propagations \textbf{(b)} Complex-time (CT) contour, combining both real- and imaginary-time propagations. This contour can only be used with the symmetrized correlation function $C(t)$.}
        \label{fig:contours}
    \end{figure}

    \subsection{Equilibrium correlation functions from process tensors}

    The forms of the equilibrium correlation functions given in Eq.~\eqref{eq:equilibrium-corr} and Eq.~\eqref{eq:symmetrized-corr} can be viewed as combinations of real-time and imaginary-time propagations and only differ in the position of the observable $B$. Taking advantage of this, we have developed a method based on propagations along the Kadanoff-Baym-like (KB) contour.
    The contour is schematically shown in Fig.~\ref{fig:contours}a, where the points $a$, $b$, and $b'$ correspond to locations where measurements of observables are made.
    Whereas $G_{AB}(t)$ requires insertions at points $a$ and $b$, the symmetrized correlator $C_{AB}(t)$ requires points $a$ and $b'$.
    
    Alternatively, the symmetrized correlator $C_{AB}(t)$ can be obtained via another approach explored by Bose~\cite{Bose2023}, based on the complex-time (CT) contour shown in Fig.~\ref{fig:contours}b, where the propagations occur over complex times. This contour differs from the KB contour in that it can only be used to compute the symmetrized correlation function for a single time $t$, though it was claimed~\cite{Bose2023} that this single calculation has low computational cost.
    We will revisit and expand upon this point in more detail.

    With a specified choice of contour, $G_{AB}(t)$ and $C_{AB}(t)$ can be Trotterized and written in terms of influence functionals $F$.
    That is, the expressions Eq.~\eqref{eq:equilibrium-corr} and Eq.~\eqref{eq:symmetrized-corr} are recast into the general form, 
    \begin{align}
        \label{eq:trotterized-corr}
        G_{AB}(t_j) = \sum_{\{s_n^{\pm}\}}  K[\{s_n^{\pm}\}] F[\{s_n^{\pm}\}].
    \end{align}

    \begin{figure*}
        \centering
        \includegraphics[width=0.8\textwidth]{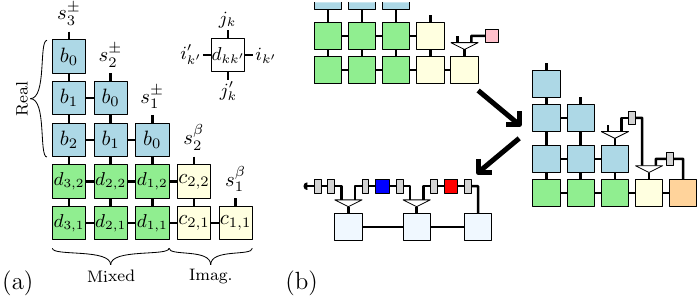}
        \caption{(Color online) (\textbf{a}) A schematic diagram of the tensor network representing the influence functionals for calculating either the thermal or symmetrized correlation functions on the KB contour. The $b_k$, $c_{k,k'}$, and $d_{k,k'}$ tensors, respectively, correlate two points on the real-time part of the contour, two points on the imaginary-time parts, and one point each on the real- and imaginary-time parts of the contour. All three- and two-legged tensors can be formed from trivial contractions over a basic four-legged tensor. (\textbf{b}) During the construction of the influence functional MPS, imaginary-time physical indices are gradually summed over as the bare system propagators (gray rectangles) are inserted. The real-time physical indices remain uncontracted as the IF construction continues with layer-by-layer MPO contraction shown in \textbf{(a)}, resulting in an MPS with $N$ open legs. Finally, the system propagators and measurements (red and blue squares) are placed in the appropriate positions with the last physical index traced over to give the desired correlation function. }
        \label{fig:KB-network}
    \end{figure*}
    Here, $n$ labels discrete positions along the contour, and $K$ encapsulates both the bare system propagations and any measurement of observables. Explicit expressions of $K$ and $F$ for thermal and symmetrized correlation functions can be found in section \ref{appendix: explicit-if} of the Supplementary Materials.
   
    In the spin-boson model, the harmonicity of the bath and its linear coupling to a diagonalizable operator of the system (with eigenvalues $s^{\pm}_k$) leads to the simple form,
    \begin{align}
        \label{eq:IF}
        F[\{s_n^{\pm}\}] &= \prod_{k=1}^N \exp \left[ - \sum_{k' = 1}^k \sum_{\sigma_1,\sigma_2=\pm} \eta^{\sigma_1\sigma_2}_{k, k'} s_{k}^{\sigma_1}s_{k'}^{\sigma_2}  \right],
    \end{align}
    where the coefficients $\eta_{k, k'}$, which we loosely refer to as temporal correlations, are dependent on the choice of  the contour.

    We formulate our approach to calculating equilibrium correlation functions on the KB contour for its capability to directly compute both $G_{AB}(t)$ and $C_{AB}(t)$.
    Specifically, this is achieved via the process tensor formulation~\cite{Pollock2018, Jorgensen2019} of the Time-Evolving Matrix Product Operator (TEMPO) algorithm, which constructs the influence functionals $F[\{s_n^{\pm}\}]$ as an MPS while maintaining the causal structure of evolution along the contour.
    This is achieved by making use of the commuting nature of all terms in $F$, which allows one to write the IF as the contraction of small tensors~\cite{Strathearn2018, Jorgensen2019} as illustrated in Fig.~\ref{fig:KB-network}a. These small tensors are defined generically as
    \begin{align}
        [d_{k, k'}]^{j_k, i_{k'} }_{j'_k, i'_{k'}}=\delta_{i_{k'},i'_{k'}} \delta_{j_k,j'_k} \exp \Big[ -\!\!\!\!\!\!\!\!\sum_{\sigma_1,\sigma_2=\pm} \eta^{\sigma_1\sigma_2}_{k, k'} s^{\sigma_1}(j_{k}) s^{\sigma_2}(i_{k'}) \Big]\hndel{.}\hnadd{,}
    \end{align}
    and similarly for $b$ and $c$. The different labels for the $b$, $c$, and $d$ tensors indicate correlations between real-real times, imaginary-imaginary times, and real-imaginary times respectively.
    After the IF has been constructed, correlation functions can be calculated by inserting measurements along the time contour, c.f.\ Eq.~\eqref{eq:trotterized-corr} and illustrated in Fig.~\ref{fig:KB-network}b.
    The correlations are extracted over the entire temporal range $\{n \delta t | n = 0, \ldots, N \}$ that the IF describes; in contrast, the CT contour limits the calculation of symmetrized correlation functions to a single time point per run, since the contour changes for every different time at which the operator $A(t)$ is measured.

    Finally, to get an idea of how to reduce computational effort in the process tensor approach, which hinges upon the bond dimension of the MPS, we now examine the behavior of temporal correlations in the IF.
    These temporal correlations, occurring between different points along the CT contour, are given by the $\eta_{k,k'}$, which are integrals of the bath correlation function, $L(t)$ (refer to Section.~\ref{appendix: explicit-if} of the Appendix for explicit expressions).
    One generally finds~\cite{Bose2023} that the magnitude of $L(t)$ is greatest along the imaginary time axis and decays with increasing distance from this axis, e.g., the points $b$ and $b'$ in Fig.~\ref{fig:contours}a should be more correlated than $b'$ and $a$, as $t_{b} - t_{b'}$ is purely imaginary while $t_{a} - t_{b'}$ acquires a large real component. In practice we find that there is large entanglement around the boundary separating the indices $\{s^{\pm}_i\}$ and $\{s^{\beta}_j\}$ and between the indices $\{s^{\beta}_j\}$ themselves, in the notation of Fig.~\ref{fig:KB-network}a.
    This issue of entanglement---and of large bond dimension by proxy---can be ameliorated by noting that the measurements required to calculate $G_{AB}$ or $C_{AB}$ will never be inserted at points on the contour between $b$ and $b'$ in Fig.~\ref{fig:contours}a.
    We can take advantage of this redundancy by using the causal nature of the construction of the influence functionals~\cite{Jorgensen2019} to sum over the unneeded indices in $\{s^{\beta}_i\}$ once they no longer have influence on the construction of the remainder of the IF.
    This process is depicted in Fig.~\ref{fig:KB-network}b. 

    \begin{figure}
        \centering
        \includegraphics[width=0.7\columnwidth]{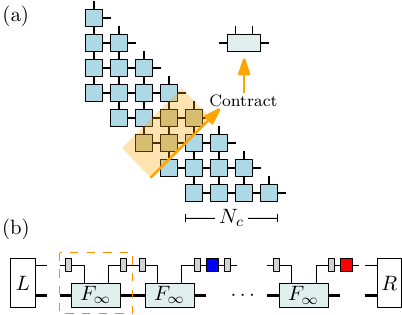}
        \caption{(Color online) Schematic depiction of how dynamics are propagated in the steady state. The element boxed in the dashed lines constitutes the transfer tensor $T_{\delta t}$ that generates system dynamics over a period $\delta t$. $L$ and $R$ are respectively the left and right eigenvectors of the transfer tensor corresponding to the largest eigenvalue. Gray boxes are system propagators over $\delta t/ 2$. Red and blue boxes are system measurements. }
        \label{fig:steady-state-mps}
    \end{figure}
    
    \subsection{Correlation functions in the steady state limit of real-time dynamics}
        The aforementioned ways of calculating $G_{AB}$ and $C_{AB}$ all construct the full equilibrium distribution $\exp(-\beta H)$, which is reflected by the need for imaginary time propagation.
        However, assuming that we are interested in the coupling of a finite sized system to a bath of infinitely many modes, one may presume the existence of a unique steady state given by the full equilibrium distribution.
        When such assumptions hold, one can convert the problem of calculating the equilibrium correlation function into the problem of calculating the steady state correlation function from an initially nonequilibrium state wherein the infinite bath is prepared at temperature $1/\beta$.

        The process of propagation to infinite times is facilitated by the observation that the influence functionals are nearly translationally invariant, except near its temporal boundaries.
        Therefore, in the infinite time/steady state limit, the influence functionals can be represented as a uniform matrix product state (uMPS), with a repeating unit cell tensor $F_{\infty}$.
        While there are multiple ways of constructing this unit cell tensor, one approach noted by Link \textit{et al.}~\cite{Link2024} is to consider the repeating strip of tensors in the TEMPO network (see highlighted rectangle in Fig. \ref{fig:steady-state-mps}a).
        Contracting these tensors using the infinite-time evolving block decimation (iTEBD) algorithm~\cite{Orus2008,Hastings2009,Unfried2023} directly yields $F_{\infty}$ (see Fig. \ref{fig:steady-state-mps}a).
        Given the nature of this construction, $F_{\infty}$ will only include temporal correlations $\eta_{|k-k'|, 0}$ for $|k-k'| = 0, \ldots, N_c$, with $N_c \delta t$ defining the memory truncation time.
        In order to accurately capture the true long-time steady state, $N_c$ must be large enough such that the temporal correlations have almost vanished.
        Thus, cases in which temporal correlations decay slowly---such as with strongly sub-Ohmic baths---may require large $N_c$ and must be treated with care~\cite{Link2024}.
        The overall cost of this approach thus depends on $N_c$ iterations of iTEBD, though we stress that this cost is not intrinsic to the steady state approach as there are different methods to obtain~$F_{\infty}$.
    
        Equipped with the uMPS $F_{\infty}$, we can now consider dynamics in the steady state.
        The $F_{\infty}$ is combined with free system propagations to construct an effective evolution tensor $T_{\delta t}$ (shown in the dashed orange box in Fig.~\ref{fig:steady-state-mps}b), describing the time evolution over a period of $\delta t$.
        The steady state is given by the right eigenvector $R$ of $T_{\delta t}$ with the eigenvalue of largest magnitude when there is a unique steady state of the dynamics.
        To $R$ there is an associated left eigenvector $L$ such that $|L\cdot R| > 0$.
        From these definitions, the steady state correlation function is obtained via the tensor network~\footnote{Alternatively, one can directly obtain an analytical approximation for $G(t)$ as a sum of exponentials using the eigendecomposition of $T_{\delta t}$, which may be useful for analytic continuation.} depicted in Fig.~\ref{fig:steady-state-mps}, corresponding to the expression,
        \begin{align}
        \label{eq:steady-state-corr}
            G(k \delta t) &= L \cdot \left( A \cdot T_{\delta t}^{k} \cdot B \right) \cdot R.
        \end{align}
        We note, however, that this steady state correlation function may not capture the correct correlation function in cases where the long-time decay of the correlation function $|G_{AB}(t) - G_{AB}(\infty)|$ is algebraic, such as found at zero temperature in the spin-boson model~\cite{Sassetti1990}.
        The fact that MPS representations of the influence functionals with fixed
        bond dimension cannot reproduce algebraically decaying correlations should be expected, as this is a direct analogue of the situation found in real-space correlation functions in the ground state of gapless 1D systems~\cite{Fannes1992}.
        Formally, to obtain a power law decay one would require that the bond dimension of $T$ grow as some polynomial of $k$, leading to an increased cost of the method.
        In practice, one can extract the asymptotic power-law behavior from the steady state approach, through a careful analysis of the spectrum of $T_{\delta t}$~\cite{Rams2018,Vanhecke2019} as a function of increasing the bond dimension of the MPS.
        
        Finally, we note that, unlike the previously stated approaches whereby the full thermal state is approximated in a systematically controllable way by the number of timesteps, the method we have outlined here for finding the steady state from real-time evolution is less straightforwardly controlled.
        In particular, the accuracy of this method will depend on the time discretization $\delta t$, the bond dimension $D$ of $F_{\infty}$, and the memory length $N_c$.
        While the first two parameters are all that control the accuracy of the methods involving complex-time propagations, the latter parameter of the memory length may play a significant role in the efficacy of the steady state approach.

    \begin{figure*}
        \centering
        \includegraphics[width=\textwidth]{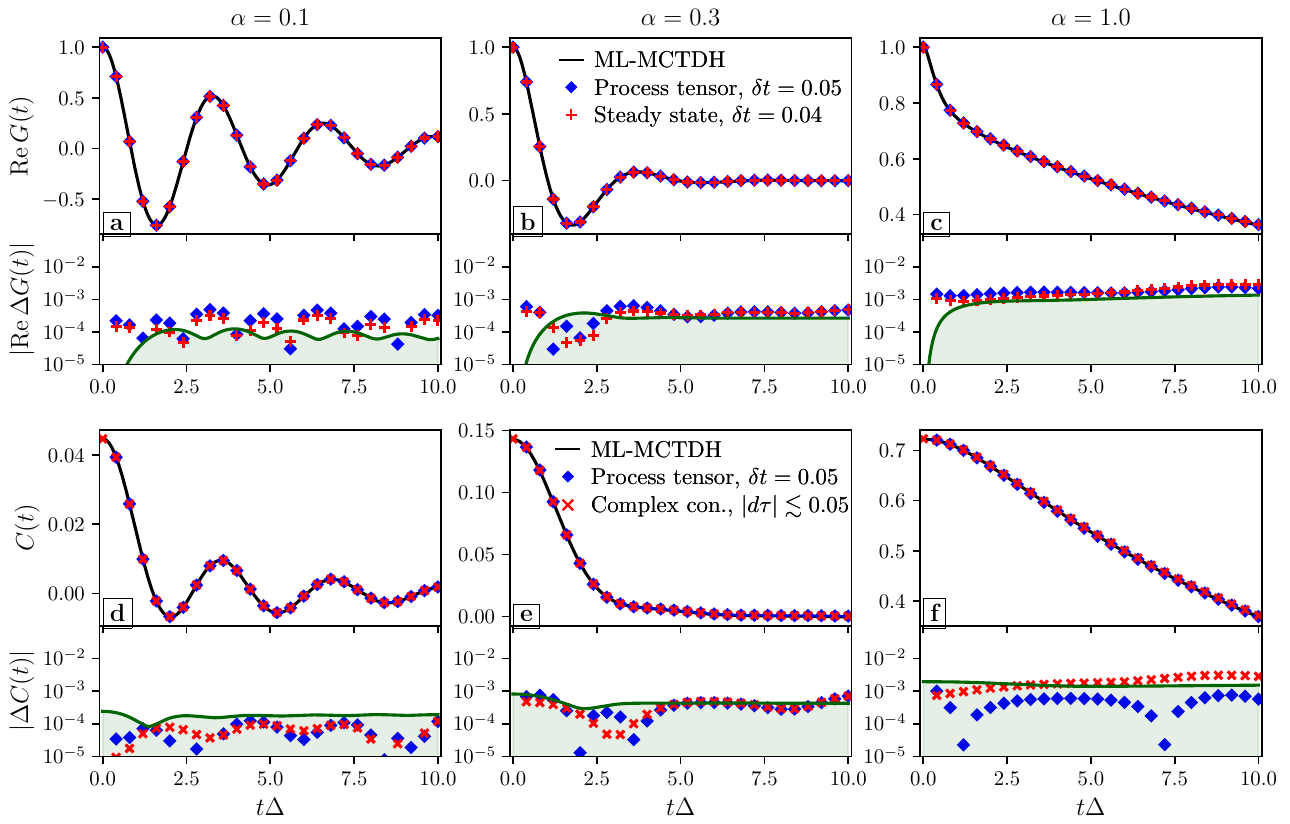}
        \caption{(Color online) Comparison of different tensor network path integral approaches to calculating the thermal correlation function $G(t)$ (upper panels) and the symmetrized correlation function $C(t)$ (lower panels), for $\omega_c = 5\Delta$ and $\beta = 5/\Delta$ for various system-bath couplings $\alpha$.
        In the bottom portion of each panel, the calculations are compared against results from ML-MCTDH, where the green shaded regions represent standard errors resulting from sampling 24576 independent trajectories from different initially pure bath states.
        The process tensor calculations (blue diamonds) were performed with a real-time discretization of $\delta t = 0.05$ and an imaginary-time discretization of $\delta \tau = 0.125$.
        In the upper panels, the steady state $G(t)$ (red plusses) is computed using a timestep of $\delta t = 0.04$ and a memory truncation time of $t_{\text{mem}} = 400$.
        In the lower panels, $C(t)$ is computed on the CT contour (red crosses), with a variable $N$ number of timesteps such that the timestep size along the contour is $\delta \tau = |t-i\beta/2|/N \approx 0.05$.}
        \label{fig:comparison}
    \end{figure*}

\section{\label{sec:results}Results}

    We use the approaches outlined to calculate autocorrelations $G_{AA}(t) \equiv G(t)$ and $C_{AA}(t) \equiv C(t)$, where $A = \spin{z}{}$, in the unbiased Ohmic spin-boson model, 
    \begin{align}
        H &= \Delta \spin{x}{} + \spin{z}{} \sum_k g_k \left( b_k + b_k^{\dagger} \right) + \sum_k \omega_k b^{\dagger}_k b_k, \label{Eq:system_bath_ham}
    \end{align}
    with spectral density defined by $J(\omega) = \sum_k g_k^2 \delta(\omega - \omega_k) = (\alpha/2) \omega \exp(-\omega/\omega_c)$ for $\omega_c = 5\Delta$.~\footnote{The normalization of $\alpha$ is such that at zero temperature in the scaling limit $\Delta/\omega_c \to 0$, the Toulouse point and the localization-delocalization transition are found respectively at $\alpha = 0.5$ and $1.0$.}
    Energy and time scales are given in units of the tunneling matrix element $\Delta = 1$, with $\hbar = k_B = 1$.
    Unless otherwise specified, the temperature of the equilibrium state is $\beta = 5/\Delta$.
    We focus our findings on this low temperature regime since it is a generally more challenging regime for tensor network influence functional methods.
    This manifests in two ways for equilibrium correlation functions: not only is more numerical effort required to propagate the dynamics in imaginary time, but also that bath correlation functions decay more slowly.

    We compare our results against calculations by the multilayer multiconfigurational time-dependent Hartree method (ML-MCTDH), in which the thermal averages are performed by sampling over initial product-state configurations of the system and bath.
    The sampling is facilitated by the use of the minimally entangled typical thermal states (METTS) algorithm~\cite{Stoudenmire2010} with the use of alternating collapse bases to reduce autocorrelation in the METTS Markov chain.~\cite{binder2017} Further details of the ML-MCTDH calculations are provided in the Supplementary Materials~\ref{appendix:MLMCTDH}.    Additionally, we compare our results for the symmetrized correlation function on the KB contour using the method described by Bose~\cite{Bose2023}, which works on the CT contour (Fig.~\ref{fig:contours}b).

    \subsection{Comparison of numerical accuracy and complexity among the three methods}
    For this low temperature ($\beta = 5/\Delta$) and fast bath ($\omega_c = 5 \Delta$) regime, we expect that the equilibrium autocorrelation function of $\spin{z}{}$ should have qualitatively similar behavior to that found in of the well understood limit of zero temperature and $\Delta/\omega_c \to 0$.
    Particularly, there should be a crossover from underdamped to overdamped to incoherent decay as the coupling strength to the bath $\alpha$ increases.
    The results for three values of the couplings ($\alpha = 0.1, 0.3, 1.0$) representative of these three regimes are shown in Fig.~\ref{fig:comparison}. %\llcomment{Might it be worth comparing the long time dynamics, e.g. we comparing the non-adiabatic rates for $\alpha=1.0$ which can be obtained from either of these correlation functions, maybe in the SI?  Given the discussion in the intro on the low frequency behaviour of $G_{AB}(t)$ and $C_{AB}(t)$?}

    In all regimes, the various methods are broadly able to reproduce the reference calculations from ML-MCTDH (Fig.~\ref{fig:comparison}).
    All calculations with the process tensor approach were performed with the same number real- and imaginary-time discretizations, $\delta t = 0.05/\Delta$ and $\delta \tau = 0.125/\Delta$ respectively.
    Where available, the steady state correlation function is computed with $\delta t = 0.04/\Delta$, with a memory cutoff $N_c = 10^5$.
    The symmetrized correlation function computed on the CT contour are all calculated with a variable discretization $N$ such that the size of the timestep along the contour $|t - i\beta/2|/N \approx 0.05$.
    The lower portions of each panel show how the results in the upper portions differ from ML-MCTDH, with statistical uncertainties depicted by the shaded green regions. Here, the statistical uncertainties are taken as $2\times$ the standard error of the mean of the $24576$ trajectories assuming independent samples.
    These errors are likely underestimated at longer times such that the apparent deviations from ML-MCTDH in Fig.~\ref{fig:comparison}(b,c) for $t \Delta \gtrsim 7$ are exaggerated.
    Additionally, we find that the errors in the process tensor approach in Fig.~\ref{fig:comparison}a are governed by the timestep discretization $\delta t$ rather than by MPS compression.
    \footnote{We show the additional convergence behaviors in Sec.~\ref{appendix:trotter} of the Supplementary Materials: the errors decrease as $O(dt^2)$ for both the correlation functions on the KB contour and for the steady state correlation function.}

    While we have established the accuracy of the various methods presented in Sec.~\ref{sec:method} for calculating the correlation functions $G(t)$ or $C(t)$, we note that the computational costs to achieve a desired level of accuracy differs greatly between the approaches.
    We quantify these differences in terms of computational complexities, memory requirements, and additional costs of the calculations, such as the computation of the $\eta_{k,k'}$ coefficients entering into the influence functionals.

\begin{figure}
        \centering
        \includegraphics[width=\columnwidth]{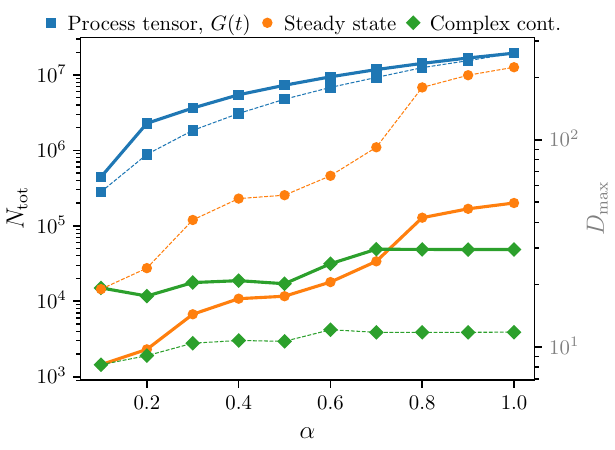}
        \caption{Maximal number $N_{\text{tot}}$ of complex elements (\textbf{solid lines}) and maximal bond dimensions $D_{\text{max}}$ (\textbf{thin dashed lines}) of the tensor network influence functionals needed to obtain results for various correlation functions (thermal correlation functions for KB-contour process tensor and steady state, and symmetrized correlation functions for CT contour process tensor) across different coupling strengths $\alpha$.}
        \label{fig:rel-tot-mem}
    \end{figure}

    We first assess the effort required to converge each method to a fixed absolute error of 0.05 for $t \Delta \in [0,10]$.
    We show in Fig.~\ref{fig:rel-tot-mem} the effort to reach this level of convergence through the maximum number $N_{\text{tot}}$ of complex-valued elements in the IF-MPS, which quantifies the memory requirement across coupling strengths $\alpha$, as well as the corresponding bond dimension $D_{\text{max}}$.
    For the CT contour, since each calculation yields $C(t)$ at a specific time point, we perform convergence tests for each time $t$ considered.
    This convergence is evaluated with respect to both the discreteness of the Trotterization and the level of compression of the MPS. The data for the CT contour reports the largest number of elements observed within the specified time range.

    In terms of computational storage, Fig.~\ref{fig:rel-tot-mem} shows that the process tensor approach is significantly more demanding than either the steady state and CT contour approaches. The process tensor also carries large bond dimensions which, when considered with the fact that the associated MPS consists of $N$ tensors as opposed to the lone unit cell tensor of the steady state's uMPS, results in overall longer runtimes for tensor operations relative to the other two approaches. It can be seen that the costs associated with the process tensor grow monotonically with increasing coupling strength $\alpha$. This trend holds similarly for the steady state approach.
    
    By contrast, the overall storage cost and bond dimensions for the CT contour approach are roughly constant across coupling strengths, though it is important to note that this is the cost associated with the calculation of $C(t)$ for a single value of $t$.
    At first glance, Fig.~\ref{fig:rel-tot-mem} may suggest that the steady state approach is a more expensive method compared to the CT contour approach. However, as noted by Link et al.~\cite{Link2024}, the bond dimension of the steady state uMPS remains relatively low during the majority of the iTEBD iterations, and only reaches the maximal bond dimension reported in Fig.~\ref{fig:rel-tot-mem} towards the end. In practice, we have found that all of the calculations in Fig.~\ref{fig:rel-tot-mem} using the steady state method took at most one minute.

    In addition to the cost of constructing and storing the influence functionals in MPS form, it is important to note that there is a cost associated with the calculation of the $\eta_{k,k'}$.
    These $\eta_{k,k'}$ are computed for every pair of points along the chosen contour.
    Hence for the process tensor approach along the KB contour (Fig.~\ref{fig:contours}a), there will be $O(MN)+O(N)+O(M)$ values of $\eta_{k,k'}$ needed if there are $2M$ steps in imaginary time and $2N$ steps in real time.
    For the calculations shown in Fig.~\ref{fig:comparison}(a,b,c), $M \sim 10$ while $N \sim 10^2$, so that about $10^3$ values of $\eta_{k,k'}$ are sufficient to compute the correlator over all $N$ time points.
    In the case of the steady state approach, for which the relevant temporal correlations enjoy a time-translationally invariant form $\eta_{k,k'} = \eta_{|k-k'|}$, only $N_{c}$ values of $\eta$'s are needed.
    This $N_{c}$ is chosen such that $\eta_{N_{c}}$ has largely decayed to zero; throughout this work $N_{c} \delta t = 400/\Delta$, from which $G(t)$ can be calculated for all $t \Delta \lesssim 400$.
    By contrast, IFs on a CT contour (Fig.~\ref{fig:contours}b) composed of $2N$ steps will require $O(N^2)$ evaluations of $\eta_{k,k'}$, one for each time point that $C(t)$ is to be evaluated over.
    We find that, in practice, this can pose significant costs for the CT contour approach when large values of $N$ are needed to converge the calculations.
    This is examined in closer detail in the following section.

    \subsection{Steady-state correlation functions versus symmetrized correlation functions}
        As previously mentioned, there are distinct drawbacks to the calculation of symmetrized correlation functions that are independent of the chosen contour or MPS representation.
        The first is that calculations along the CT contour become increasingly difficult as the temperature is lowered or as the time increases.
        This is due to the need to maintain a certain level of discretization of the contour in order to control the Trotter error, which may be compounded by the need to calculate $C(t)$ to longer times as $C$ decays more slowly with increasing $\alpha$ or lower temperature.
        Second, in view of Eq.~\eqref{eq:relation}, features in the spectra obtained from $C(t)$ may become exponentially suppressed with inverse temperature $\beta$ relative to what may be found from $G(t)$.
        In such cases, errors may be exponentially amplified when attempting to recover $G(t)$ from $C(t)$.
        It would be crucial therefore to ensure that the calculation of $C(t)$ is properly converged, particularly with respect to Trotter errors.        
        In this work, we leave aside questions of the numerical stability of converting $C(t)$ to $\widetilde{G}(\omega)$, and hence will not dwell on this second point.
        
        \begin{figure}
            \centering
            \includegraphics[width=\columnwidth]{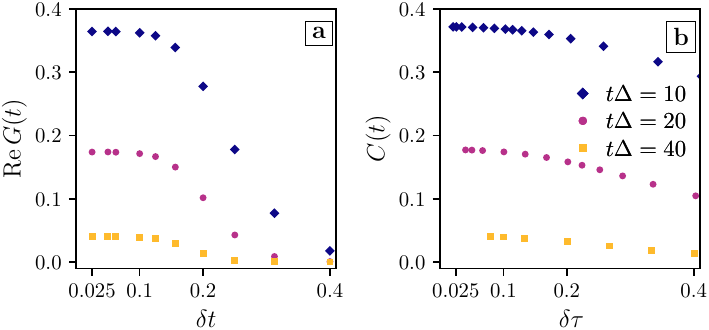}
            \caption{Convergence of the (\textbf{a}) steady state and (\textbf{b}) symmetrized correlation functions with respect to discretization ($\delta t$ or $\delta \tau = |t - i\beta/2|/N$) at three times, for strong coupling ($\alpha = 1.0$) to a low temperature ($\beta \Delta = 5$) Ohmic bath. The left-most points in (\textbf{b}) correspond to $N = 500$.}
            \label{fig:long-strong}
        \end{figure}
        
        We will now focus on the first point, examining the low temperature regime and comparing the use of the steady-state approach to calculate $G(t)$, versus the calculation of $C(t)$ on the CT contour. For numerical illustration, we examine more closely the long time behavior of $G(t)$ and $C(t)$ in Fig.~\ref{fig:comparison}(c,f).
        To resolve the decay behavior at intermediate times, we look at the correlation functions at times $t \Delta = 10, 20, 40$ to examine the effort required to converge calculations with respect to the discretization in real ($\delta t$) or complex time ($\delta \tau = |t - i \beta/2|/N$ for number of timesteps $N$ along one leg of the contour in Fig.~\ref{fig:contours}b).
        This is shown in Fig.~\ref{fig:long-strong}.
        Even though the convergence behaviors differ between the steady state and CT contour approaches, they are converged at roughly similar sizes of Trotter steps along their respective contours.
        This highlights a potential pitfall of the CT contour approach, that the accurate resolution of $C(t)$ may become more difficult with increasing $t$.
        If the correlation function decays as $f(t)$ and the overall Trotter error scales as $O(t^2/N^2)$, then in order to have fixed level of relative error at time $t$ would require $N \sim O(t f(t)^{-1/2})$.
        In the intermediate time regime $t \lesssim \beta$, it is expected~\cite{Sassetti1990} that $f \sim t^{-2}$.
        Thus the total number of $\eta_{k,k'}$ computations per time $t$ would scale as $O(t^4)$, which would make the CT contour approach much more computationally expensive than Fig.~\ref{fig:rel-tot-mem} might suggest.
        
    \subsection{Steady-state correlation functions at zero temperature}
        \begin{figure}
            \centering
            \includegraphics[width=\columnwidth]{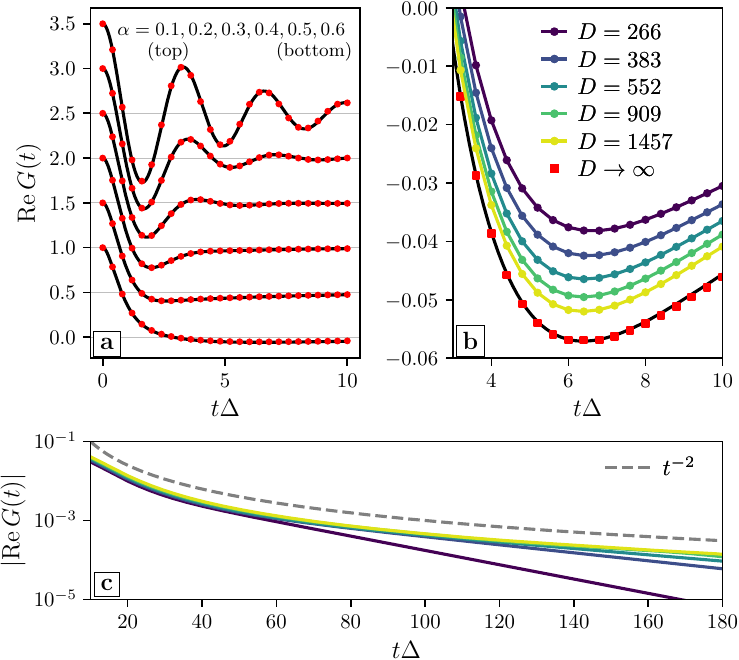}
            \caption{(Color online) The real part of zero temperature correlation functions $G(t) = \Theta(t) \langle \spin{z}{}(t) \spin{z}{} (0) \rangle$ computed using the steady state influence functionals approach with $\delta t \Delta = 0.04$ (points) in comparison with converged ML-MCTDH calculations (black lines). (\textbf{a}) $\Re G(t)$ across coupling strengths $\alpha = 0.1, 0.2, 0.3, 0.4, 0.5, 0.6$, from top to bottom, for an Ohmic bath with frequency cutoff scale $\omega_c = 5 \Delta$. The data is shifted for clarity, with the respective zeros shown in light gray lines. (\textbf{b}) Closeup of $\Re G(t)$ for $\alpha = 0.6$, showing the convergence of the steady state approach to the ML-MCTDH result with increasing bond dimension $D$. 
            Red points denote an extrapolation to the $D = \infty$ limit by means of a power law fit.
            (\textbf{c}) Convergence of the long-time decay of $|\text{Re } G(t)|$ towards the expected $\sim t^{-2}$ behavior (grey dashed line), with increasing bond dimension, for $\alpha = 0.6$.
            }
            \label{fig:ss-zeroT}
        \end{figure}
    
        Finally, we examine the utility of the steady state approach for calculating equilibrium correlation functions by focusing on the regime that cannot be exactly calculated by methods relying on imaginary time propagation.
        We will focus on the zero temperature limit of the spin-boson model, at the regime of couplings below the putative localization transition ($\alpha_c \simeq 1.25$~\cite{Strathearn2018} for $\omega_c = 5\Delta$), comparing the results against direct ground state simulations using ML-MCTDH.

        In Fig.~\ref{fig:ss-zeroT}a, we show the zero temperature results for $\Re G(t)$ across coupling strengths from top to bottom, $\alpha = 0.1, 0.2, 0.3, 0.4, 0.5, 0.6$, which have been shifted for clarity. 
        The results largely agree well with ML-MCTDH calculations, though closer inspection at larger couplings $\alpha \sim 0.6$ shows slight deviations from the reference results (Fig.~\ref{fig:ss-zeroT}b).
        It can be seen that the steady state correlation functions are converging towards the ML-MCTDH calculations as the bond dimension increases.
        Additionally, given that the zero temperature correlation functions in the unbiased spin-boson model are expected to decay algebraically~\cite{Sassetti1990}, it is reasonable to anticipate the departure of the steady state calculations---which approximate $G(t)$ as the sum of exponentials---from the true result due to finite bond dimensions~\cite{Vidal2007, Zauner2015}.
        This behavior can be explicitly seen in Fig.~\ref{fig:ss-zeroT}c, where the long-time behavior of $G(t)$ is exponential for small bond dimensions, e.g.\ $D = 266$, and converges to the expected $\sim t^{-2}$ decay as $D$ increases.
        Lastly, a previous investigation by Rams et al.~\cite{Rams2018} suggests some utility of extrapolating to the infinite bond dimension limit of infinite MPS's using a power-law ansatz, $x_D \sim x_{\infty} + c D^p$, even away from criticality.
        Applying this ansatz to our approximations to the correlation function, we recover reasonably good agreement with ML-MCTDH (Fig.~\ref{fig:ss-zeroT}b).

\section{\label{sec:discussion}Discussion}
    
    Focusing on tensor network influence functionals, we have compared three different approaches to calculating equilibrium correlation functions in open quantum systems in which the system is linearly coupled to a continuous bosonic bath using (i) influence functionals defined on an imaginary- and real-time Kadanoff-Baym contour, to obtain $G_{AB}(t)$ (ii) influence functionals defined on a complex time contour to obtain $C_{AB}(t)$ and (iii) the extraction of the steady state correlation using a fixed point method for the influence functionals defined only on the real-time axis.
    For these three methods, we have considered the effort, as quantified through the complexities of different algorithms and their associated memory requirements, to converge calculations to a given level of relative error in the correlation function.
    In the case of the symmetrized correlation $C_{AB}(t)$, we stress that this consideration does not take into account the effort required to recover the equilibrium correlation function $G_{AB}(t)$ from $C_{AB}(t)$.
    For the purposes of this work, we shall consider this recovery procedure to be a separate mathematical problem.

    We first discuss the process tensor method associated with the KB contour. As originally motivated, the process tensor method constructs an IF-MPS containing information on all two-point correlation functions---as well as \textit{all possible} multi-point ones---that are of interest. This is in contrast to the IF-MPS constructed along the CT contour\hndel{in Fig.~\ref{fig:contours}b}, for which only \textit{one} two-point correlation is of physical relevance. Thus one must preserve more information in an IF-MPS on the KB contour than on the CT contour, a fact which allows methods in the latter case to keep computational costs low by tracing out unneeded indices to reduce the overall size of the MPS. 
    In light of this, we note that one can take a similar philosophy towards the IF-MPS on the KB contour, which we expect would ameliorate the costs that we have observed (Fig.~\ref{fig:rel-tot-mem}).
    In this work, however, we have focused on the advantage that the KB contour IF-MPS can be reused for other simulations as long as the details of the bath and maximum simulation time remain the same. 
    We generally find that this benefit comes at the cost of higher runtime and storage costs.
    
    Unlike the process tensor method, the CT contour requires matrix product states of much lower bond dimension and computational storage among the three methods (Fig.~\ref{fig:rel-tot-mem}).
    This, when combined with the fact that the MPS associated with the CT contour is defined over $O(N)$ sites, leads to comparable memory requirements to the steady state approach, which has a uMPS defined by a single unit cell tensor of significantly larger bond dimension. Despite the associated lower costs of tensor operations, this method has a few disadvantages, the foremost of which is that it can only obtain $C_{AB}$ for a single time.
    While the method is trivially parallelizable, this concern can become troublesome if one requires $C_{AB}$ for many points in time. A consequence of the need to reconstruct an IF-MPS for each value of $t$ is the effort required to compute all the associated temporal correlations along the CT contour, $\eta_{k,k'}$, for each time $t$. We empirically find this to be a non-negligible cost to the running time of the method.
    This cost is enhanced especially when dealing with slowly decaying correlation functions, which subsequently require larger $N$ at long times to converge the Trotter errors.

    Finally, we discuss the steady state method, specifically that constructed from the iTEBD algorithm. Similar to the process tensor method, the steady state method can directly compute all possible multi-point correlation functions in a single run. However, it does not encounter similar runtime or memory costs as it only needs to operate on a single tensor as opposed to $O(N)$ tensors. In addition, the steady state approach, being formulated on the real-time axis, does not suffer from the problem of needing to recompute many $\eta_{|k-k'|}$. And unlike either the CT contour or the KB contour approaches for calculating correlation functions, since the steady state method is not strictly formulated on a finite time interval, it has the advantage of being able to compute correlation functions to arbitrarily large $t$ provided that $F_{\infty}$ contains temporal correlations $\eta_{|k-k'|, 0}$ where $t \lesssim |k-k'| \delta t$. Lastly, since the steady state approach explicitly does not contain imaginary-time propagation, it can be used in cases of extremely low temperatures.
    This is particularly the case when the correlation functions decay rapidly. In the case of algebraically decaying correlations, 
    the uMPS approach becomes less efficient for large $t$. Other types of tensor network representations of the influence functionals that naturally capture such correlations should then be explored.

    We anticipate that the conclusions we have reached here about the most efficient ways of using tensor network influence functionals to obtain equilibrium correlation functions should not be specific for the spin-boson model, and should hold similarly for fermionic impurity models as well.

    \noindent\textit{Note:} During the final stages of this work, we became aware of similar studies by Guo and Chen~\cite{Guo2024,Chen2024} that calculate equilibrium and steady state correlation functions for fermionic impurity problems.

    {\noindent}{\bf Supplementary Material}:
    See the supplementary materials for 1) More data on the convergence behavior of the thermal correlation function on the KB contour and the steady state correlation function; 2) explicit expressions for the $\eta_{k,k'}$ and system propagators on the KB and CT contours; and 3) details of the ML-MCTDH approach used for evaluating symmetrized and equilibrium correlation functions.

    {\noindent}{\bf Code availability}: The code is available on Github at \url{https://github.com/nguye66h/sb-ecfs}.

    {\noindent}{\bf Acknowledgements}:
    This work was performed with support from the U.S.\ Department of Energy, Office of Science, Office of Advanced Scientific Computing Research, Scientific Discovery through Advanced Computing (SciDAC) program, under Award No. DE-SC0022088.
    This research used resources of the National Energy Research Scientific Computing Center (NERSC), a U.S. Department of Energy Office of Science User Facility located at Lawrence Berkeley National Laboratory, operated under Contract No. DE-AC02-05CH11231.
    The Flatiron Institute is a division of the Simons Foundation.
    L.P.L.\ acknowledges the support of the Engineering and Physical Sciences
Research Council [grant EP/Y005090/1].

    {\noindent}{\bf Author Declarations}:

    {\noindent}{\bf Conflict of Interest}:
    The authors have no conflicts to disclose.

    % https://credit.niso.org/
    {\noindent}{\bf Author Contributions}:
    \textbf{Haimi Nguyen}: Conceptualization (equal); Data curation (equal); Investigation (equal); Methodology (equal); Software (equal); Visualization (equal); Writing - original draft preparation (equal); Writing - reviewing \& editing (equal).
    \textbf{Nathan Ng}: Conceptualization (equal); Data curation (equal); Investigation (equal); Methodology (equal); Software (equal); Visualization (equal); Writing - original draft preparation (equal); Writing - reviewing \& editing (equal).
    \textbf{Lachlan Lindoy}: Conceptualization (equal); Data curation (equal); Methodology (equal); Investigation (equal); Validation (lead); Writing - original draft preparation (equal); Writing - reviewing \& editing (equal).
    \textbf{Gunhee Park}: Conceptualization (equal); Writing - reviewing \& editing (equal).
    \textbf{Andrew Millis}: Conceptualization (equal); Funding acquisition (equal); Supervision (supporting); Writing - reviewing \& editing (equal).
    \textbf{Garnet Chan}: Conceptualization (equal); Funding acquisition (equal); Supervision (supporting); Writing - reviewing \& editing (equal).
    \textbf{David Reichman}: Conceptualization (equal); Funding acquisition (equal); Project administration (lead); Supervision (lead); Writing - original draft preparation (equal); Writing - reviewing \& editing (equal).

    {\noindent}{\bf Data Availability}:
    The data that support the findings of this study are available upon reasonable request.

\bibliographystyle{aipnum4-2}
\bibliography{references}

\clearpage
\newpage
\setcounter{page}{1}
\setcounter{equation}{0}
\setcounter{section}{0}
\onecolumngrid

\appendix
\renewcommand{\thefigure}{ (SM \arabic{figure})}
\setcounter{figure}{0}

\begin{center}
\Large Supplementary Materials for ``Correlation Functions From Tensor Network Influence Functionals: The Case of the Spin-Boson Model''
\end{center}

\vspace*{2em}

This supplement is divided into 3 parts:
\begin{itemize}
    \item Scaling of Trotter errors
    \item Expressions for $\eta_{k, k'}$ along various contours
    \item Details of the multilayer multiconfigurational time-dependent Hartree method (ML-MCTDH) calculations
    % \item Construction of the process tensor 
\end{itemize}

\section{Scaling of Errors}
\label{appendix:trotter}
In this section, we examine convergence behaviors of our numerical methods to validate their stability and accuracy. Apart from the compression of the matrix product state, the only other source of error comes from discretizing the evolution operators, i.e. Trotter errors. Given that we converge with respect to the MPS compression, we should expect to see the theoretical Trotter error scaling. 

\begin{figure}[h]
    \centering
    \includegraphics[width=1.0\textwidth]{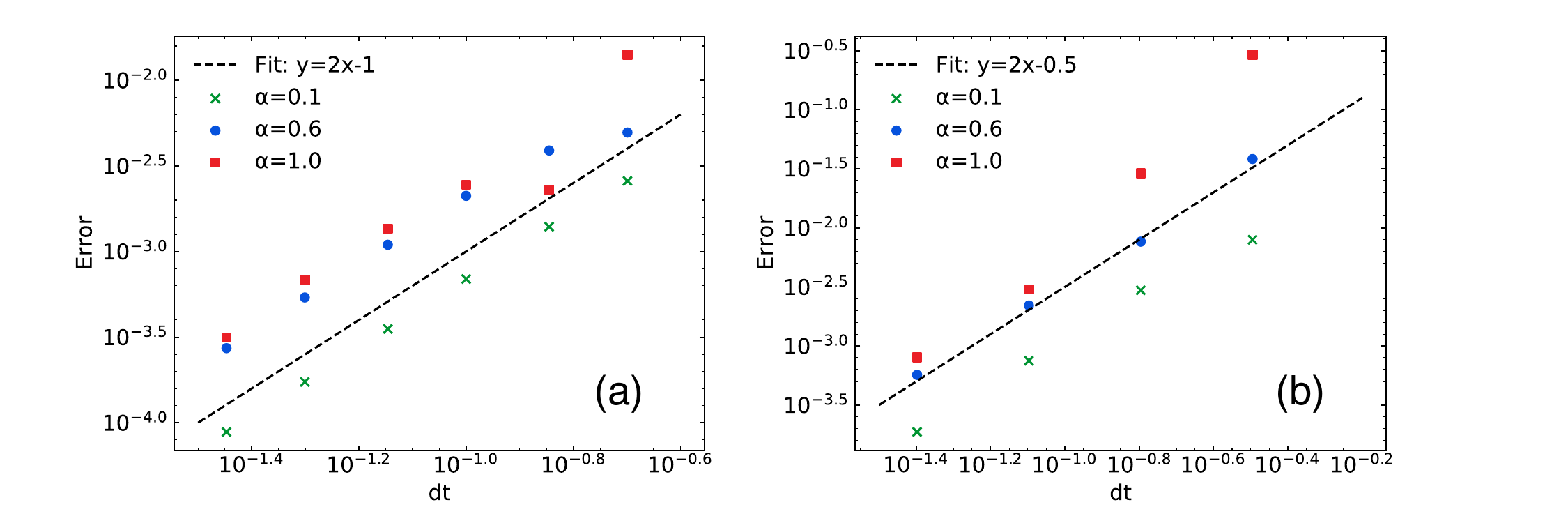}
    \caption{ \textbf{(a)} The convergence behavior of the thermal correlation computed on the KB contour for various system-bath couplings strengths $\alpha$, as quantified by the absolute difference at $t=1\Delta$, $|G_{dt} - G_{dt/2}|$, for two timesteps $dt$ and $\frac{dt}{2}$. The dashed line is a guide to the eye, showing the $dt^2$ convergence 
    \textbf{(b)} The convergence behavior of the steady state correlation function at $t=10 \Delta$. The dashed line again corresponds to $dt^2$ scaling.}
    \label{fig:trotter}
\end{figure}

In our process tensor method on the Kadanoff-Baym-like (KB) contour, we implement the symmetric Trotter splitting, where the error per step is of  $O(dt^3)$. We compute the thermal correlation functions at a fixed time of $1\Delta$, resulting in $N \sim O(\frac{1}{dt})$. Subsequently, the cumulative error at the fixed time point is proportional to $O(dt^2)$. Similarly, the construction of the unit cell $F_\infty$ also makes use of the symmetric Trotter decomposition with local error of $O(dt^3)$. Keeping the memory cutoff constant, the cumulative error should scale with $O(dt^2)$. In Figure \ref{fig:trotter}, we show exactly these relationships for a few coupling strengths within the same unbiased spin-boson model as in the main paper. 

\section{Explicit Expressions of Thermal and Symmetrized Correlation Functions} 
\label{appendix: explicit-if}
\begin{figure}[h]
    \centering
    \includegraphics{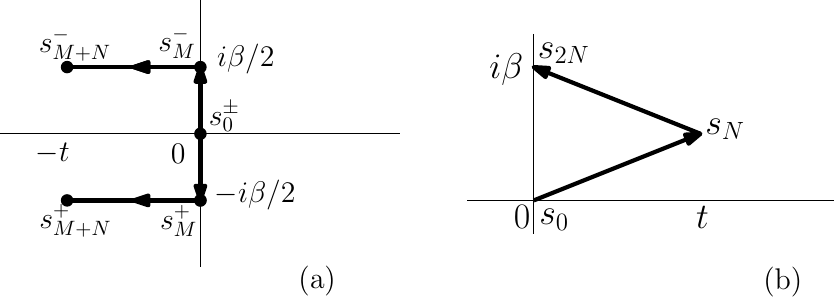}
    \caption{Schematic depiction of the time points on \textbf{(a)} the Kadanoff-Baym-like (KB) contour and \textbf{(b)} the complex-time (CT) contour.}
    \label{fig:SM-contours}
\end{figure}

Here we detail explicitly the terms in Eq.~\ref{eq:trotterized-corr} in the main text, reproduced here for convenience:
\begin{align}
        \bar{C}_{AB}(t_j) = \sum_{\{s_n^{\pm}\}}  K[\{s_n^{\pm}\}] F[\{s_n^{\pm}\}],
\end{align}
where $\bar{C}_{AB}(t_j)$ denotes thermal or symmetrized correlation function at time $t_j$.
The KB contour is divided into the purely imaginary-time propagation (over $M$ steps) and purely real-time propagation (over $N$ steps).
The system states $\{s_n^{\pm}\}$ and their corresponding locations on the contour are shown in Fig.~\ref{fig:SM-contours}

\subsection{Kernel $K$}
We define the discretized timesteps used in the following sections as
\begin{align}
    \Delta t=&\frac{t}{N},\\
    \Delta\beta=&\frac{\beta}{2M},\\
    \Delta\tau=&\frac{t-i\beta/2}{N}.
\end{align}
\subsubsection{KB Contour Thermal Correlation Function:}
\begin{align}
    K[\{s^\pm\}]=&\langle s^-_{M+N} | s^+_{M+N} \rangle \langle s^+_{M+N} | B e^{iH_{\text{sys}}\Delta t/2} | s^+_{M+N-1} \rangle \langle s^+_{M+N-1}| e^{iH_{\text{sys}}\Delta t} |s^+_{M+N-2} \rangle \dots \langle s^+_{M+1} | e^{iH_{\text{sys}}\Delta t} | s^+_{M} \rangle \\ \nonumber
    &\langle s^+_{M} | e^{iH_{\text{sys}}\Delta t/2} e^{-H_{\text{sys}}\Delta \beta/2} | s^+_{M-1} \rangle \langle s^+_{M-1} | e^{-H_{\text{sys}}\Delta \beta} | s^+_{M-2} \rangle \dots \langle s^+_{1} | e^{-H_{\text{sys}}\Delta \beta} | s^+_{0} \rangle \\ \nonumber
    &\langle s^+_{0} | e^{-H_{\text{sys}}\Delta \beta} | s^-_{0} \rangle \langle s^-_{0} | e^{-H_{\text{sys}}\Delta \beta} |s^-_{1} \rangle \dots \langle s^-_{M-2} | e^{-H_{\text{sys}}\Delta \beta} | s^-_{M-1} \rangle \\ \nonumber
    &\langle s^-_{M-1} | e^{-H_{\text{sys}}\Delta \beta/2} A e^{-iH_{\text{sys}}\Delta t/2} |  s^-_{M} \rangle \langle s^-_{M} | e^{-iH_{\text{sys}}\Delta t } | s^-_{M+1} \rangle \dots \\ \nonumber
    &\langle s^-_{M+N-2} | e^{-iH_{\text{sys}}\Delta t} | s^-_{M+N-1} \rangle \langle s^-_{M+N-1} | e^{-iH_{\text{sys}}\Delta t/2} | s^-_{M+N} \rangle.
\end{align}
\subsubsection{KB Contour Symmetrized Correlation Function:}
\begin{align}
    K[\{s^\pm\}]=&\langle s^-_{M+N} | s^+_{M+N} \rangle \langle s^+_{M+N} | B e^{iH_{\text{sys}}\Delta t/2} | s^+_{M+N-1} \rangle \langle s^+_{M+N-1}| e^{iH_{\text{sys}}\Delta t} |s^+_{M+N-2} \rangle \dots \langle s^+_{M+1} | e^{iH_{\text{sys}}\Delta t} | s^+_{M} \rangle \\ \nonumber
    &\langle s^+_{M} | e^{iH_{\text{sys}}\Delta t/2} e^{-H_{\text{sys}}\Delta \beta/2} | s^+_{M-1} \rangle \langle s^+_{M-1} | e^{-H_{\text{sys}}\Delta \beta} | s^+_{M-2} \rangle \dots \langle s^+_{1} | e^{-H_{\text{sys}}\Delta \beta} | s^+_{0} \rangle \\ \nonumber
    &\langle s^+_{0} | e^{-H_{\text{sys}}\Delta \beta/2} A e^{-H_{\text{sys}}\Delta \beta/2} | s^-_{0} \rangle \langle s^-_{0} | e^{-H_{\text{sys}}\Delta \beta} |s^-_{1} \rangle \dots \langle s^-_{M-2} | e^{-H_{\text{sys}}\Delta \beta} | s^-_{M-1} \rangle \\ \nonumber
    &\langle s^-_{M-1} | e^{-H_{\text{sys}}\Delta \beta/2} e^{-iH_{\text{sys}}\Delta t/2} |  s^-_{M} \rangle \langle s^-_{M} | e^{-iH_{\text{sys}}\Delta t } | s^-_{M+1} \rangle \dots \\ \nonumber
    &\langle s^-_{M+N-2} | e^{-iH_{\text{sys}}\Delta t} | s^-_{M+N-1} \rangle \langle s^-_{M+N-1} | e^{-iH_{\text{sys}}\Delta t/2} | s^-_{M+N} \rangle .
\end{align}

\subsubsection{CT Contour Symmetrized Correlation Function:}
\begin{align}
    K[s^\pm]=&\langle s_0 | B e^{iH_{\text{sys}} \Delta \tau^*/2} | s_1 \rangle \langle s_1 | e^{iH_{\text{sys}} \Delta \tau^*} | s_2 \rangle \\ \nonumber
    & \dots \langle s_{N-1} | e^{iH_{\text{sys}} \Delta \tau^*} | s_N \rangle \langle s_N | e^{iH_{\text{sys}} \Delta \tau^*/2} A e^{-iH_{\text{sys}} \Delta \tau/2} | s_{N+1} \rangle \langle s_{N+1} | e^{-iH_{\text{sys}} \Delta \tau} | s_{N+2} \rangle \dots \\ \nonumber
    & \langle s_{2N-1} | e^{-iH_{\text{sys}} \Delta \tau} | s_{2N} \rangle \langle s_{2N} | e^{-iH_{\text{sys}} \Delta \tau/2} | s_{0} \rangle .
\end{align}

\subsection{The Influence Functional}

We begin this section by defining the bath correlation function as follows:
\begin{align}
    L(t)=\int_{0}^{\infty} d\omega J(\omega) \left[ \coth \left( \frac{\beta\omega}{2} \right) \cos \left( \omega t \right) - i \sin \left( \omega t \right) \right].
\end{align}

\subsubsection{KB Contour:}
For the KB contour, the influence functional tensor has the following form:
\begin{align}
    F[\{s_n^{\pm}\}] &= \prod_{k=0}^{N+M-1} \exp \left[ - \sum_{k' = 0}^{k} \eta^{++}({k, k'}) s_k^+ s_{k'}^+ + \eta^{+-}({k, k'}) s_k^+ s_{k'}^- + \eta^{-+}({k, k'}) s_k^- s_{k'}^+ + \eta^{--}({k, k'}) s_k^- s_{k'}^- \right],
\end{align}
where for $k \neq k'$,
\begin{align}
    \eta^{++}({k, k'})=\left[ \eta^{--}({k, k'})\right ]^*=\int_{t_{s_k^+}}^{t_{s^+_{k+1}}} dt_1 \int_{t_{s_{k'}^+}}^{t_{s^+_{k'+1}}}  dt_2 L (t_1-t_2) ,
\end{align}
\begin{align}
    \eta^{+-}({k, k'})=\left[ \eta^{-+}({k, k'})\right] ^*=\int_{t_{s_k^+}}^{t_{s^+_{k+1}}} dt_1 \int_{t_{s_{k'+1}^-}}^{t_{s^-_{k'}}}  dt_2 L(t_1-t_2),
\end{align}
and for $k = k'$,
\begin{align}
    \eta^{++}({k, k})=\left[ \eta^{--}({k, k})\right] ^*= \int_{t_{s_k^+}}^{t_{s^+_{k+1}}} dt_1 \int_{t_{s_{k}^+}}^{t_1}  dt_2 L (t_1-t_2),
\end{align}
\begin{align}
    \eta^{+-}({k, k})=\eta^{-+}({k, k})=\frac{1}{2} \int_{t_{s_k^+}}^{t_{s^+_{k+1}}} dt_1 \int_{t_{s_{k+1}^-}}^{t_{s^-_{k}}}  dt_2 L(t_1-t_2).
\end{align}
The contour in Fig.  \ref{fig:SM-contours}a defines the meaning of $t_{s_k}$ in our  expressions of influence functional phase. More explicitly,\\
for $k \leq M$
\begin{align}
    t_{s_k^+}&=\frac{-i\beta}{2M}k, \\
    t_{s_k^-}&=\frac{i\beta}{2M}k, 
\end{align}
and for $k > M$
\begin{align}
    t_{s_k^+}&=\frac{-i\beta }{2} - \frac{t}{N}(k-M),\\
    t_{s_k^-}&=\frac{i\beta }{2} - \frac{t}{N}(k-M).
\end{align}

\subsubsection{CT Contour:}

The influence functional of the complex-time (CT) contour takes a much simpler form:
\begin{align}
    F[\{s_n\}] &= \prod_{k=1}^{2N} \exp \left[ - \sum_{k'=1}^{k} \eta({k, k'}) s_k s_{k'} \right],
\end{align}
where
\begin{align}
    \eta({k, k'})= \int_{t_{s_{k'}}}^{t_{s_{k'-1}}} dt_1 \int_{t_{s_k}}^{t_{s_{k-1}}}  dt_2 L (t_1-t_2),
\end{align}
\begin{align}
    \eta({k, k})=\frac{1}{2}\int_{t_{s_k}}^{t_{s_{k-1}}} dt_1 \int_{t_{s_k}}^{t_{s_{k-1}}}  dt_2 L (t_1-t_2).
\end{align}
As in the previous section, we provide the contour in Fig.  \ref{fig:SM-contours}b to define the time points $t_{s_k}$ in the expressions of influence functional.
% \begin{align}
%     d\tau = \frac{t+i\beta/2}{N}
% \end{align}
% \begin{align*}
%     t_{s_k}&=k\frac{t+i\beta/2}{N} & \text{For k < N} \\
%     t_{s_k}&=t+i\beta/2 - (k-N)\frac{t-i\beta/2}{N} & \text{For k > N} \\
% \end{align*}

% \begin{figure}[h]
%     \centering
%     \includegraphics{figures/contour_complex.pdf}
%     \caption{Schematic depiction of the time points on the complex-time contour.}
%     \label{fig:complex}
% \end{figure}
% \section{Detailed description of the Process-Tensor Algorithm}
% \label{pt-algo}
% \begin{figure}[h]
%     \centering
%     \includegraphics{figures/pt.pdf}
%     \caption{Schematic depiction of the process-tensor influence functional construction. $d_{k,k'}$, $c_{k,k'}$, and $b_{\Delta k}$ are correlations between real-imaginary, imaginary-imaginary, and real-real times respectively. All the physical indices of the green squares, which represent the imaginary times, are to be summed over. The bright blue square represents measurements and imaginary-time system propagation.}
%     \label{fig:pt-if-construct}
% \end{figure}
% We provide a schematic depiction of the construction of process-tensor influence functional in Fig \ref{fig:pt-if-construct}. The physical indices of the green squares represent imaginary times and therefore unnecessary to calculate the correlation functions. They are to be summed over. 

\section{Details of the ML-MCTDH Calculations \label{appendix:MLMCTDH}} 
In this section, we provide details of the multilayer multiconfigurational time-dependent Hartree method (ML-MCTDH) calculations presented in the main text. 
Specifically, we present the bath discretization strategy and the evolution scheme employed to evolve the ML-MCTDH wavefunctions, and outline the use of the minimally entangled typical thermal state (METTS) algorithm for the evaluation of the correlation functions.

The ML-MCTDH approach makes use of a tree tensor network based ansatz for the wavefunction of a closed quantum system.  Here, in order to treat the dynamics of the spin-boson models considered in the main text, we consider a system described by the discrete system-bath Hamiltonian given in Eq. \ref{Eq:system_bath_ham}.  
Consequently, it is necessary to construct a discretized representation of the continuum ohmic spectral density $J(\omega) = \sum_k g_k^2 \delta(\omega-\omega_k) = (\alpha/2) \omega \exp(-\omega/\omega_c)$. 
To do this we have used the bath-spectral-density-orthogonal polynomial based quadrature scheme described in Ref.~\onlinecite{deVega2015}, in which the frequencies and coupling constants of the bath modes are obtained from an $N$-point Gaussian quadrature rule constructed from the orthogonal polynomials that are orthogonal with respect to the weight function $W(\omega) = J(\omega) \Theta(\omega_{max} - \omega)$.
The effect of bath modes with $\omega_k > \omega_{max}$ is approximated by first performing a polaron transform before tracing out the high frequency modes.  This process gives rise to a renormalized tunneling matrix element 
\begin{equation}
\widetilde{\Delta} = \Delta \exp \left[-2 \int_{\omega_{max}}^\infty \frac{J(w)}{\omega^2} \coth(\beta \omega/2) \mathrm{d}\omega \right],
\end{equation} 
and allows for more rapid convergence than simply discarding high frequency modes.
In all calculations presented, the number of quadrature points, $N$, and maximum frequency cutoff, $\omega_{max}$, are convergence parameters.  For the parameter regimes considered in the main text, $N=256$ and $\omega_{max}=50\Delta$ were found to provide sufficiently converged dynamics.

Several strategies exist for evolving the coefficients in the tree tensor network wavefunction central to the ML-MCTDH method.  Here we have used the projector splitting integrator\cite{LUBICH2015,KieriSJNA2016,KLOSS2017,BONFANTI2018252,Lindoy2021, CERUTI2021}, which avoids numerical stability issues associated with direct product initial conditions in the standard ML-MCTDH approach\cite{Lindoy2021b}.  In all calculations, a single-site evolution algorithm was used, augmented with a subspace expansion approach similar to those proposed in Ref. \onlinecite{MendiveJCP2020} for the standard ML-MCTDH algorithm to allow for growth of the bond-dimension (or number of single-particle functions) throughout the dynamics.

For the evaluation of zero temperature correlation functions, ground states were obtained using the single-site tree tensor network state optimization algorithm described in Ref.~\onlinecite{LarssonJCP2019}, however, with the additional use of subspace expansion allowing for adaptive control of bond dimension throughout the optimization.  

\subsection{Evaluation of Finite Temperature Correlation Functions with ML-MCTDH}
Several possible strategies have been proposed for the use of tensor network wavefunctions for the evaluation of thermal properties of quantum systems \cite{wang2006, white2009,Stoudenmire2010, craig2011, binder2017}.  Here we apply the minimally entangled typical thermal state (METTS) algorithm\cite{white2009, Stoudenmire2010, binder2017}, for the evaluation of equilibrium and symmetrized correlation functions using ML-MCTDH.  The idea behind the METTS algorithm is the expansion of the Boltzmann operator as a sum over projectors onto METTS states, $\ket{\phi(i)}$, as
\begin{equation}
    e^{-\beta \hat{H}} = \sum_{i} P(i) \ket{\phi(i)} \bra{\phi(i)},
\end{equation}
with 
\begin{equation}
    \ket{\phi(i)}= P(i)^{-1/2} e^{-\beta\hat{H}/2} \ket{i}, \label{eq:metts}
\end{equation}
 $P(i) = \bra{i}e^{-\beta\hat{H}}\ket{i}$, and $\ket{i}$ is an orthonormal, product basis state in the full system and bath Hilbert space.  A Markov chain of METTS states can be generated efficiently through a two step process: \cite{Stoudenmire2010}
 \begin{enumerate}
     \item Starting from a product basis state $\ket{i}$, construct the METTS $\ket{\phi(i)}$ according to Eq.~\ref{eq:metts}
     \item Collapse the basis state onto a product state $\ket{j}$ with probability $P(i\rightarrow j) = \|\braket{j}{\phi(i)}\|^2$ using a projective measurement.
 \end{enumerate}
 The transition probabilities between METTS states satisfy detailed balance\cite{Stoudenmire2010, binder2017} and can be used to evaluate thermal expectation values and correlation functions by averaging properties over the Markov chain of METTS states.   
\subsubsection{Evaluation of Correlation Functions}
Within the METTS scheme, the equilibrium correlation functions of the form
\begin{align}
    G_{AB}(t) &= \Theta(t)\frac{1}{Z}\mathrm{tr}\left[e^{-\beta\hat{H}} e^{-i\hat{H}t} \hat{A} e^{i\hat{H}t} \hat{B}   \right] \\
    &= \Theta(t)\frac{1}{Z}\sum_{i} P(i) \bra{\phi(i)} \hat{A} e^{i\hat{H}t} \hat{B} e^{-i\hat{H}t} \ket{\phi(i)}
\end{align}
can be evaluated as 
\begin{equation}
G_{AB}(t) = \Theta(t)\frac{1}{Z}\sum_{i} P(i) \bra{\phi(i)} \hat{A} e^{i\hat{H}t} \hat{B} e^{-i\hat{H}t} \ket{\phi(i)}.
\end{equation}
That is, for each sampled METTS, $\ket{\phi(i)}$, we time evolve the states $\ket{\phi(i)}$ and $\hat{A}^\dagger\ket{\phi(i)}$ and evaluate the matrix elements of $\hat{B}$ between these two states.

Similarly, the symmetrized correlation functions
\begin{equation}
    C_{AB}(t) = \Theta(t)\frac{1}{Z}\mathrm{tr}\left[ e^{-i\hat{H}t} e^{-\beta\hat{H}/2} \hat{A} e^{-\beta\hat{H}/2} e^{i\hat{H}t} \hat{B}   \right]
\end{equation}
can be computed as
\begin{equation}
    C_{AB}(t) = \Theta(t)\frac{1}{Z}\sum_{i} P(i) \bra{\phi_{\hat{A}}(i)} e^{i\hat{H}t} \hat{B} e^{-i\hat{H}t} \ket{\phi(i)},
\end{equation}
where 

\begin{equation}
\ket{\phi_{\hat{A}}(i)}= P(i)^{-1/2} e^{-\beta\hat{H}/2} \hat{A}^\dagger\ket{i}.
\end{equation}
As such, for each METTS sample $\ket{\phi(i)}$, we need to evolve the states $\ket{\phi(i)}$ and $\ket{\phi_{\hat{A}}(i)}$ and evaluate the matrix elements of $\hat{B}$ between these two states, to evaluate samples for the symmetrized correlation functions.  As the time evolved METTS state $e^{-i\hat{H}t} \ket{\phi(i)}$ is common to the evaluation of both symmetrized and thermal correlation functions, we can compute both functions by evolving only three wavefunctions per METTS sample. 

\subsubsection{METTS Collapse Bases}
The METTS approach can suffer from significant correlation between elements in the Markov chain when using a single collapse basis\cite{binder2017}.  These autocorrelations can be reduced through the use of multiple collapse bases.  Here we employ two alternating collapse bases: the first is the bosonic number operator basis, the second is a basis that allows for more efficient mixing between different non-interacting boson operator states.
For a specific boson mode, this basis is defined by a transformation matrix $U$ that is obtained by constructing the nearest unitary to the matrix with elements:
\begin{equation}
M_{ik} = e^{-E_{ik}(\omega)^2/\sigma(\omega)^2} e^{i\theta_{ik}}, 
\end{equation}
where $\theta_{ik} \in [0, 2\pi)$ is a random phase variable, and $E_{ik}(\omega) = \omega (i-k)$ is the energy difference between the two states assuming a non-interacting bosonic Hamiltonian.  Here we define 
\begin{equation}
    \sigma(\omega) = \begin{cases}
        \kappa \omega & \kappa \omega < \epsilon \\
        \epsilon & \mathrm{otherwise} \\
    \end{cases}
\end{equation}
This choice of basis function allows for mixing of a state $i$ with states nearby in energy.

\end{document}